\documentclass{ws-ijmpcs}
\begin{document}

\markboth{Syaefudin Jaelani, for the ALICE Collaboration}
{Nuclear modification factor and elliptic flow in Pb--Pb collisions}

%
\catchline{}{}{}{}{}
%

\title{Measurement of the D-meson nuclear modification factor and elliptic flow in Pb--Pb collisions at $\sqrt{s_{\text{\tiny{NN}}}}$ = 5.02 TeV with ALICE at the LHC}

\author{Syaefudin Jaelani, for the ALICE Collaboration\footnote{SYAEFUDIN JAELANI}}

\address{Physics Department, Utrecht University, Princetonplein 1\\
Utrecht, 3584 CC,
Netherlands\footnote{Physics Department, Utrecht University}\\
syaefudin.jaelani@cern.ch}



\maketitle

\begin{history}
\published{Day Month Year}
\end{history}

\begin{abstract}

Heavy-flavour hadrons are effective probes to study the Quark-Gluon Plasma (QGP) formed in ultra-relativistic heavy-ion collisions. The ALICE Collaboration measured the D-mesons (D$^{0}$, D$^{+}$, D*$^{+}$ and D$^{+}_{\text s}$) production in Pb--Pb collisions at $\sqrt{s_{\text{\tiny{NN}}}}$  = 5.02 TeV. The in-medium energy loss can be studied by means of the nuclear modification factor ($R_{\text{AA}}$). The comparison between the D$^{+}_{\text s}$  and the non-strange D-meson $R_{\text{AA}}$ can help to study the hadronisation mechanism of the charm quark in the QGP. In semi-central collisions the measurement of the D-meson elliptic flow, $v_{2}$, at low $p_{\text{T}}$ allows to investigate the participation of the heavy quarks in the collective expansion of the system while at high $p_{\text{T}}$ it constrains the path-length dependence of the energy loss. Furthermore the Event-Shape Engineering (ESE) technique is used to measure D-meson elliptic flow in order to study the coupling of the charm quarks to the light quarks of the underlying medium.

\end{abstract}

\section{Introduction}	

Heavy flavours are effective probes to study the Quark-Gluon Plasma (QGP) which is formed in ultra-relativistic heavy-ion collisions. Due to their large masses, heavy quarks are produced in hard scattering processes in the early stages of the collisions before the QGP formation time, which is about 0.3-1.5 fm/$c$ at Large Hadron Collider (LHC) energies \cite{fmliu}. Thus, they experience the full evolution of the medium and lose part of their energy interacting with its constituents via inelastic (gluon radiation) \cite{gyu}$^{,}$\cite{bai} or elastic scatterings (collisional processes) \cite{thoma}\cdash\cite{brat2}. The in-medium energy loss can be studied by measuring the nuclear modification factor ($R_{\text{AA}}$), which compares the $p_{\text{T}}$-differential yield in Pb--Pb collisions with that in pp collisions, $R_{\text{AA}}(p_\text{T}) = ({\text{d}} N_{\text{AA}} / {\text{d} p_{\text T}} ) / (\left\langle T_{\text {AA}} \right\rangle \cdot \text{d} \sigma_{\text{pp}} / \text{d} p_{\text T})$. The comparison of the non-strange and strange D mesons can provide information about the mechanism of charm hadron formation \cite{kuz}. In addition, the information on the transport properties of the medium is obtained by the measurement of the azimuthal anisotropy in the momentum distribution of heavy-flavour hadrons, by the elliptic flow $v_{2} = \left\langle \cos (2(\varphi - \Psi_{2})) \right\rangle $. The measurement of the elliptic flow $v_{2}$ of D-meson production in semi-central collisions allows to investigate, at low $p_{\text{T}}$, the participation of the heavy quarks in the collective expansion of the system and their thermalization in the medium. In addition at high $p_{\text{T}}$, the measurement of $v_{2}$ can provide information on the path length dependence of parton energy loss \cite{bats}$^{,}$\cite{gyu2}.

\section{D-meson reconstruction}

In ALICE D mesons are reconstructed in the hadronic decay channels D$^{0} \rightarrow$ K$^{-} \pi^{+}$ (with branching ratio, BR, of 3.93 $\pm$ 0.04\%), D$^{+}\rightarrow$ K$^{-} \pi^{+} \pi^{+}$ (BR of 9.46 $\pm$ 0.24\%), D$^{*+} \rightarrow$ D$^{0} \pi^{+}$ (BR of 67.7 $\pm$ 0.5\%) and D$^{+}_{\text s}$ $\rightarrow$  $\phi \pi^{+} \rightarrow$ K$^{-}$K$^{+}\pi^{+}$ (BR of 2.27 $\pm$ 0.08\%) \cite{colla}, and their charge conjugates.
The decay topologies are resolved via secondary vertex-reconstruction thanks to the excellent performance of the Inner Tracking System. Background is reduced by applying topological selections in order to maximize the signal-to-background ratio. In addition, particle identification is carried out for charged pions and kaons, with the Time Projection Chamber and the Time Of Flight detector. Finally, an invariant mass analysis is used to extract the D-meson yields. The efficiency and acceptance corrections are obtained from MC simulations based on HIJING \cite{wang} and PYTHIA 6 \cite{sjo} event generators. The prompt yield of D mesons is obtained by subtracting from the inclusive yield the secondary yield from beauty hadrons as estimated based on FONLL calculations \cite{cac}$^{,}$\cite{alice}. The V0 scintillators, that cover the pseudorapidity region --3.7  $< \eta <$ --1.7 and 2.8 $< \eta <$ 5.1, provide centrality and event plane angle (estimator of $\Psi_{2}$).

\section{Prompt D-meson nuclear modification factor and elliptic flow}
\begin{figure}[pb]
\centerline{
\includegraphics[width=5.2cm]{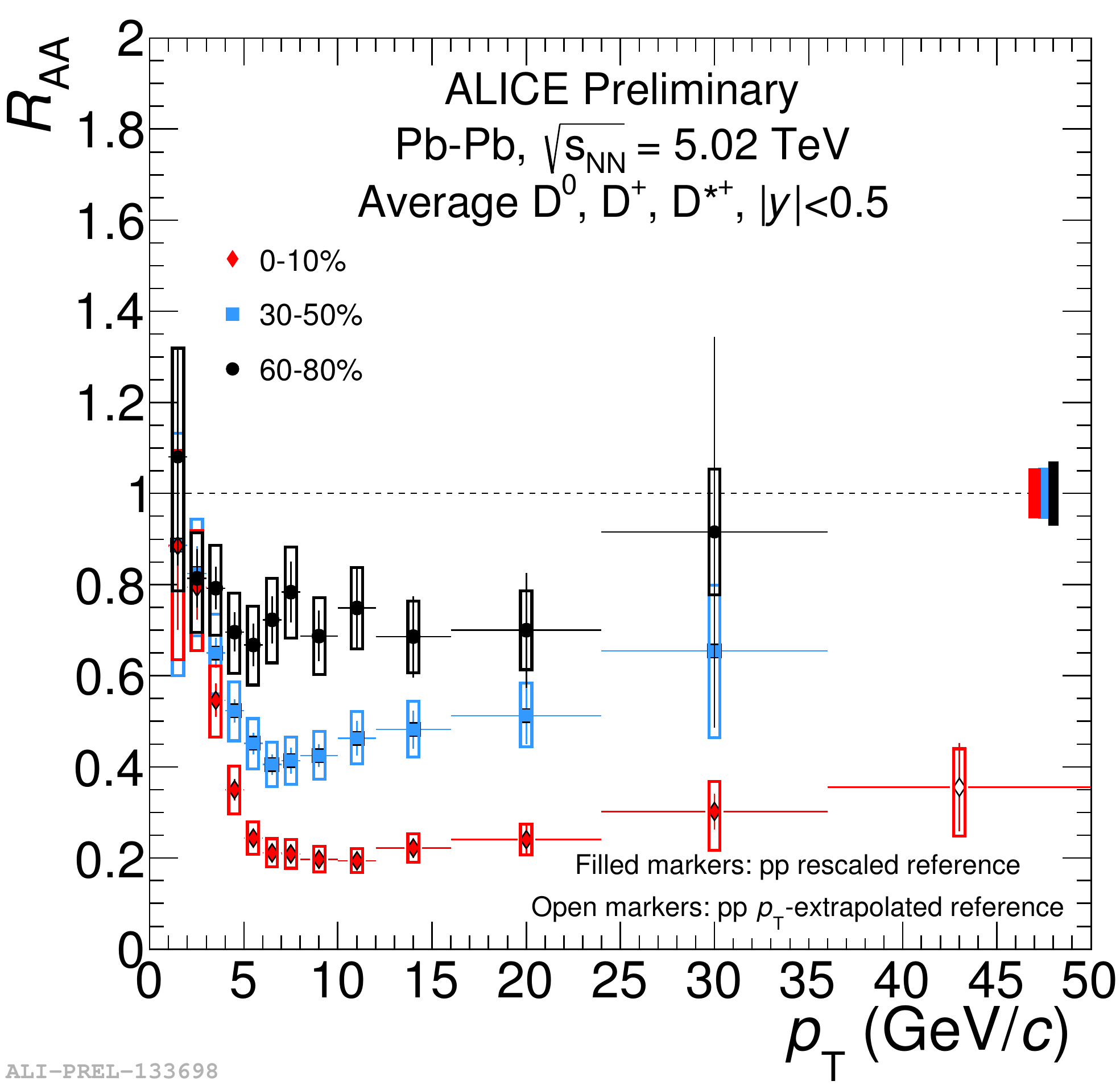} \hspace{0.5cm}
\includegraphics[width=5.2cm]{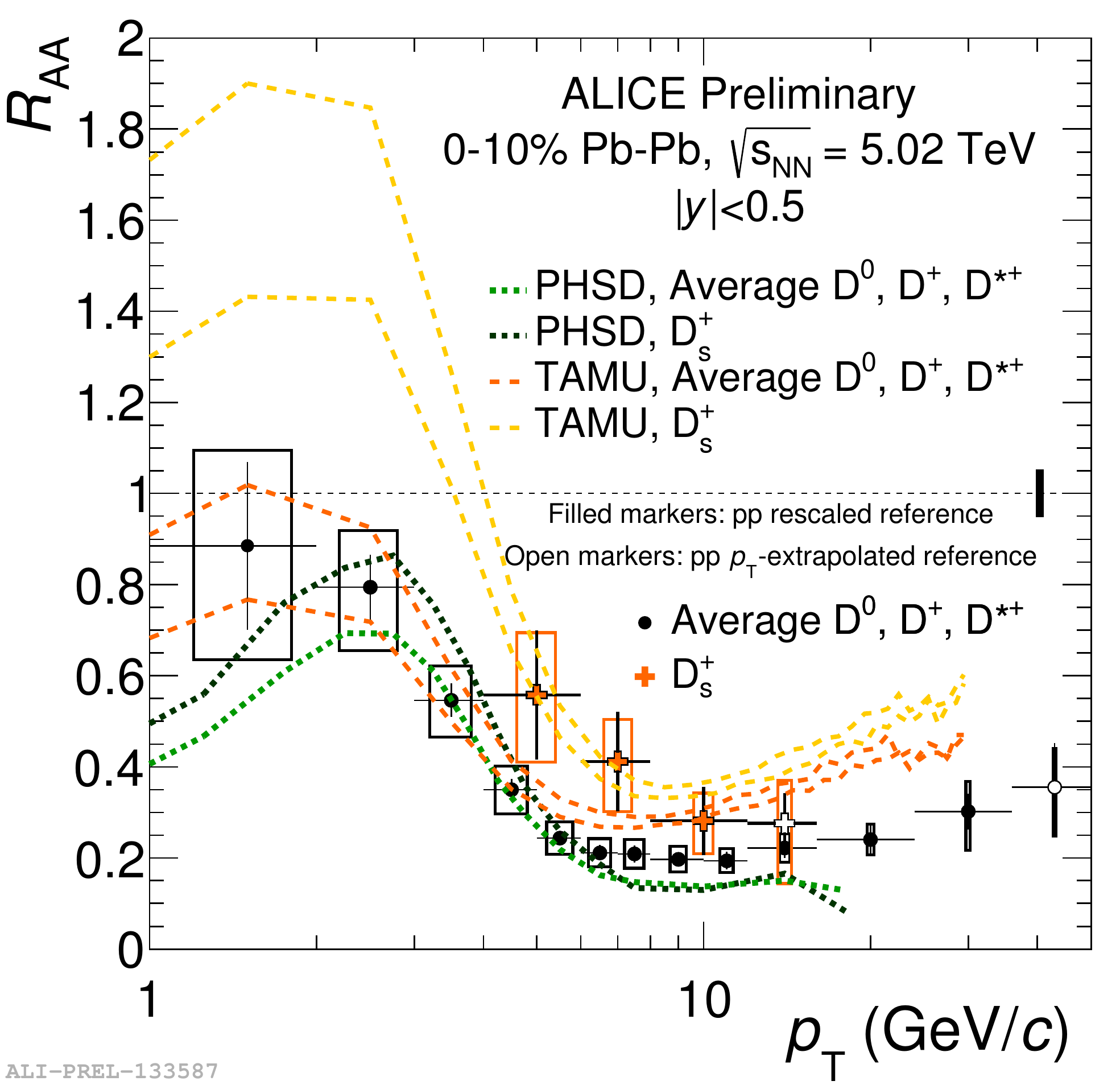}
}
\vspace*{8pt}
\caption{Left: average $R_{\text{AA}}$ of prompt D$^{0}$, D$^{+}$ and D*$^{+}$ in central 0--10\% (red), semi-central 30--50\% (blue) and peripheral 60--80\% (black) in Pb--Pb collisions at $\sqrt{s_{\text{\tiny{NN}}}}$  = 5.02 TeV $^{15}$. Right: $R_{\text{AA}}$ of prompt D$^{+}_{\text s}$ compared with the average non-strange D-meson $R_{\text{AA}}$  in Pb--Pb collisions at $\sqrt{s_{\text{\tiny{NN}}}}$  = 5.02 TeV in 0--10\% centrality class $^{15}$. The D$^{+}_{\text s}$ model predictions includes hadronisation mechanism of charm hadrons via coalescence in the QGP $^{17,18}$. 
\label{Fig1}}
\end{figure}

The $R_{\text{AA}}$ of prompt D$^{0}$, D$^{+}$, D*$^{+}$ and D$^{+}_{\text s}$ is measured in three different centrality classes (central 0--10\%, semi central 30--50\% and peripheral 60--80\%) in Pb--Pb collisions at $\sqrt{s_{\text{\tiny{NN}}}}$  = 5.02 TeV \cite{pas}. The proton-proton reference was obtained by scaling the measurement cross section at $\sqrt{s_{\text{\tiny{NN}}}}$  = 7 TeV \cite{alice2} to $\sqrt{s_{\text{\tiny{NN}}}}$  = 5.02 TeV using the FONLL prediction. Fig. \ref{Fig1}, in the left panel, shows the increasing suppression of the average non-strange D-meson $R_{\text{AA}}$ from peripheral to central collisions up to a factor about 5 for $p_{\text{T}} >$ 5 GeV/$c$. These results are compatible within uncertainties with Run 1 data at $\sqrt{s_{\text{\tiny{NN}}}}$  = 2.76 TeV \cite{pas}, showing at the same time an improvement in terms of precision and $p_{\text{T}}$ reach. The central value of the prompt D$^{+}_{\text s}$ $R_{\text{AA}}$ is higher relative to non-strange D-meson for all centralities, however, due to the large statistical and systematic uncertainties, additional data are needed to conclude. Fig. \ref{Fig1}, on the right panel, shows the comparison of non-strange D-meson and D$^{+}_{\text s}$ $R_{\text{AA}}$ in the 0--10\% centrality class with model predictions that include hadronisation mechanism of charm via coalescence \cite{he}$^{,}$\cite{song}.

\begin{figure}[pb]
\centerline{
\includegraphics[width=6.5cm]{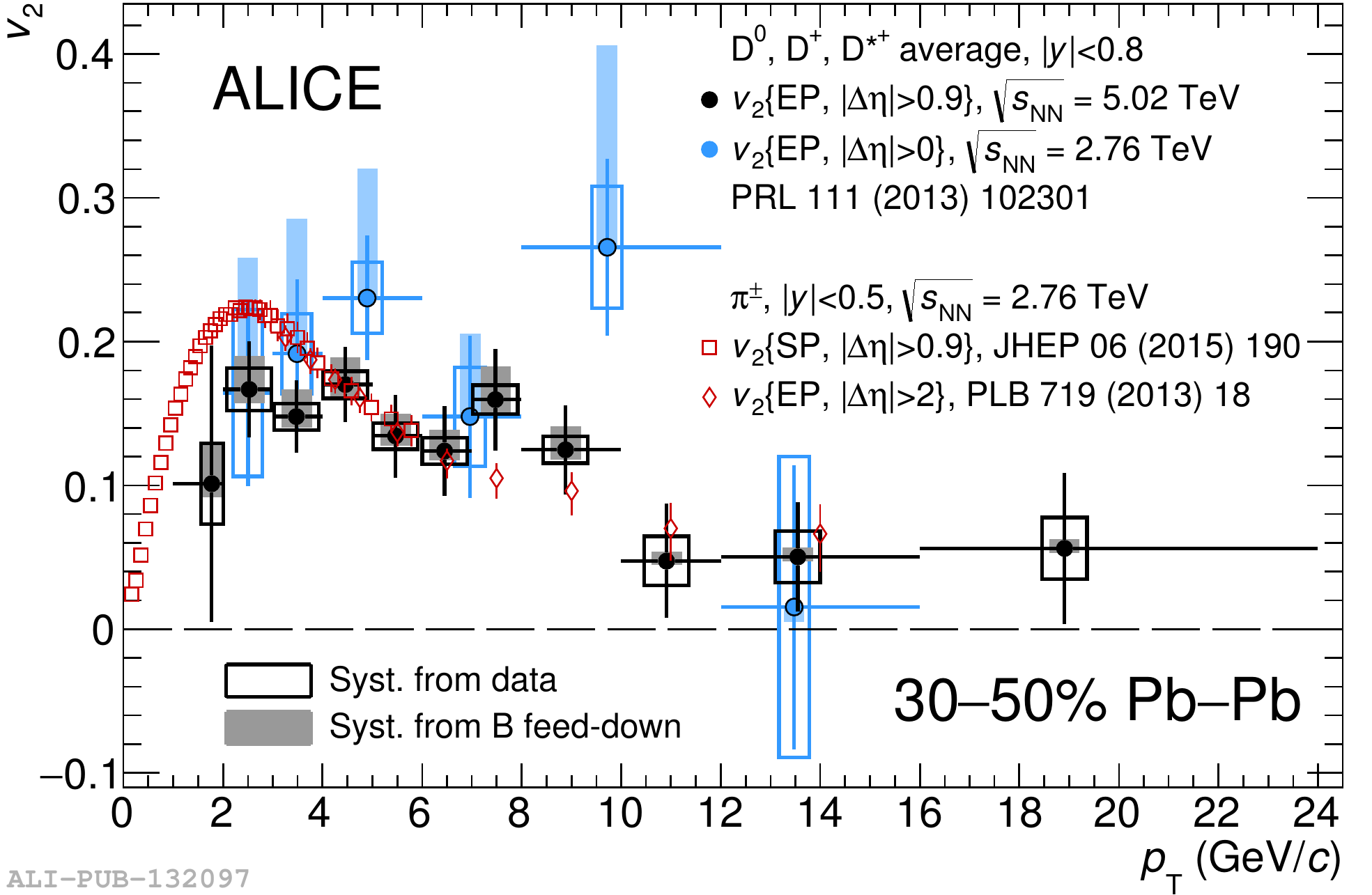} \hspace{0.3cm}
\includegraphics[width=5.2cm]{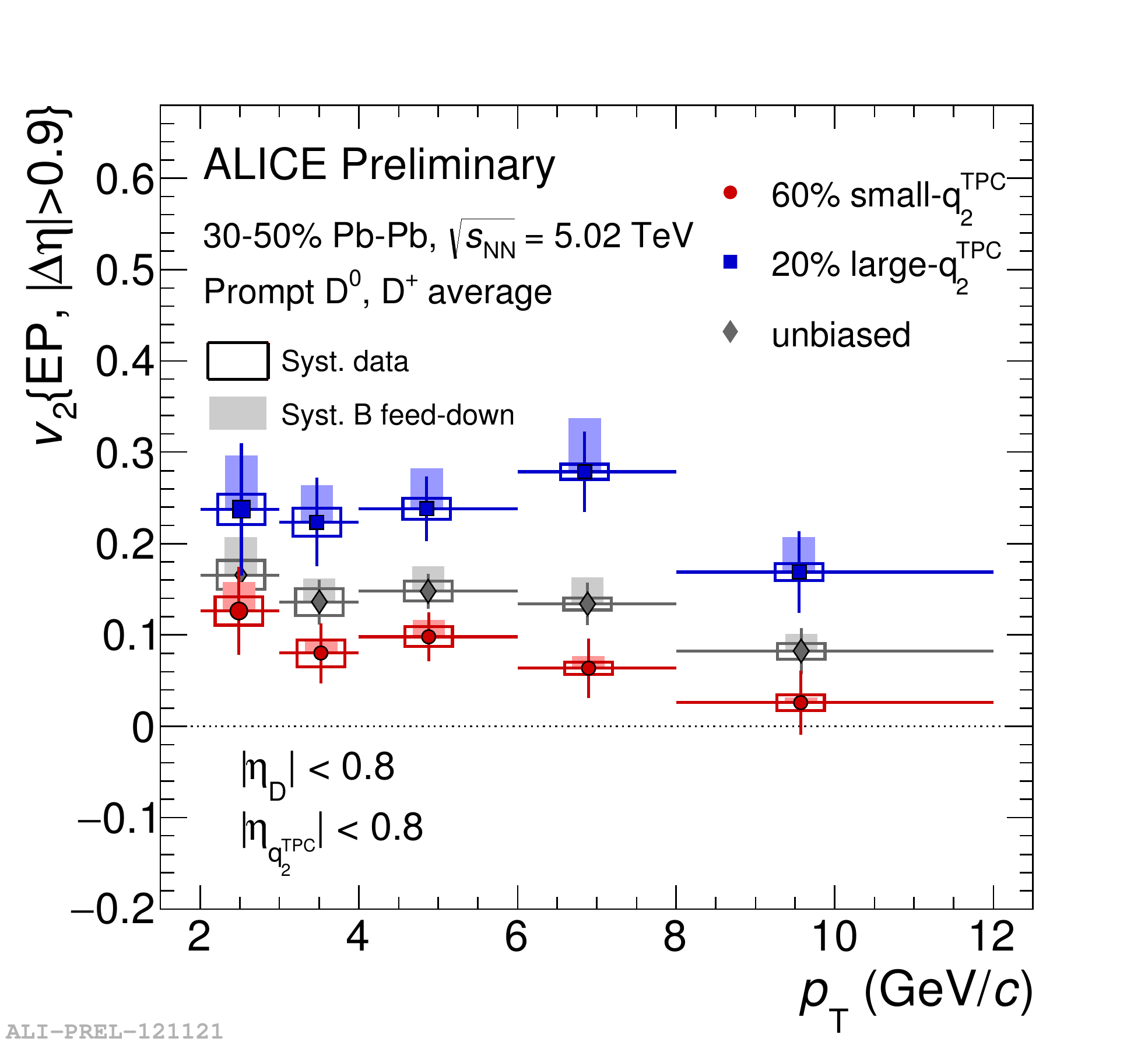}
}
\vspace*{8pt}
\caption{Left: average non-strange D-meson elliptic flow $v_{2}$ as a function of $p_{\text T}$ in Pb--Pb collisions at $\sqrt{s_{\text{\tiny{NN}}}}$  = 5.02 TeV in 30--50\% centrality class (black), compared with the same measurement of $v_{2}$ (blue) and with the $\pi^{\pm} \ v_{2}$ at $\sqrt{s_{\text{\tiny{NN}}}}$  = 2.76 TeV (red) $^{19}$. Right: average D$^{0}$ and D$^{+} \ v_{2}$ in semi-central 30--50\%  Pb--Pb collisions at $\sqrt{s_{\text{\tiny{NN}}}}$  = 5.02 TeV for the 60\% of events with smallest $q_{2}$ and the 20\% of events with largest $q_{2}$ compared to the unbiased result.
\label{Fig2}}
\end{figure}

The measurement of prompt D-meson elliptic flow $v_{2}$ in Pb--Pb collisions at $\sqrt{s_{\text{\tiny{NN}}}}$  = 5.02 TeV in the 30-50\% centrality class \cite{alice3} is reported in Fig. \ref{Fig2}, on the right panel. The elliptic flow coefficients, $v_{2}$, of prompt D$^{0}$, D$^{+}$ and D*$^{+}$ are compatible with each other within uncertainties and confirm a non-zero D-meson $v_{2}$ for 2 $<p_{\text{T}} <$  10 GeV/$c$, indicating that charm quarks participate in the collective expansion dynamics. The measurement of the D-meson $v_{2}$ at $\sqrt{s_{\text{\tiny{NN}}}}$ = 5.02 TeV is compatible with the same measurement at $\sqrt{s_{\text{\tiny{NN}}}}$ = 2.76 TeV and with the charged pion $v_{2}$ measured in the same centrality class (see Fig. \ref{Fig2}, left panel). The D$^{+}_{\text s}$ $v_{2}$ hints a non-zero value in 2 $<p_{\text{T}} <$ 8 GeV/$c$ with significance of about 2.6 $\sigma$ and it is also compatible within uncertainties to that of non-strange D mesons. The Event-Shape Engineering technique was used  for the first time to investigate the D$^{0}$ and D$^{+}$ $v_{2}$ in 30--50\% centrality class. The second-harmonic reduced flow vector, $q_{2} = |\mathit{Q}_{2}| / \sqrt{\mathit{M}}$, can be used to quantify the eccentricity (average $v_{2}$) of the events, where $\mathit{M}$ is the multiplicity and $\mathit{Q}_{2}$ is the second-harmonic flow vector. The events were divided into two groups, namely 60\% of events with smallest $q_{2}$ and 20\% of events with largest $q_{2}$. The average D-meson $v_{2}$ for the two $q_{2}$ values compared to the unbiased are presented on the right panel of Fig. \ref{Fig2}. The separation between the D-meson $v_{2}$ in the large $q_{2}$ and small $q_{2}$ hints to sensivity of charm quarks to the light quarks collectivity and to the event-by-event fluctuations in the initial state.  Though, the effect can be increased by autocorrelations and non-flow correlations between $q_{2}$ and D-meson $v_{2}$, since they are measured in the same pseudorapidity region $|\eta| < 0.8$.

\begin{figure}[pb]
\centerline{
\includegraphics[width=4.5cm]{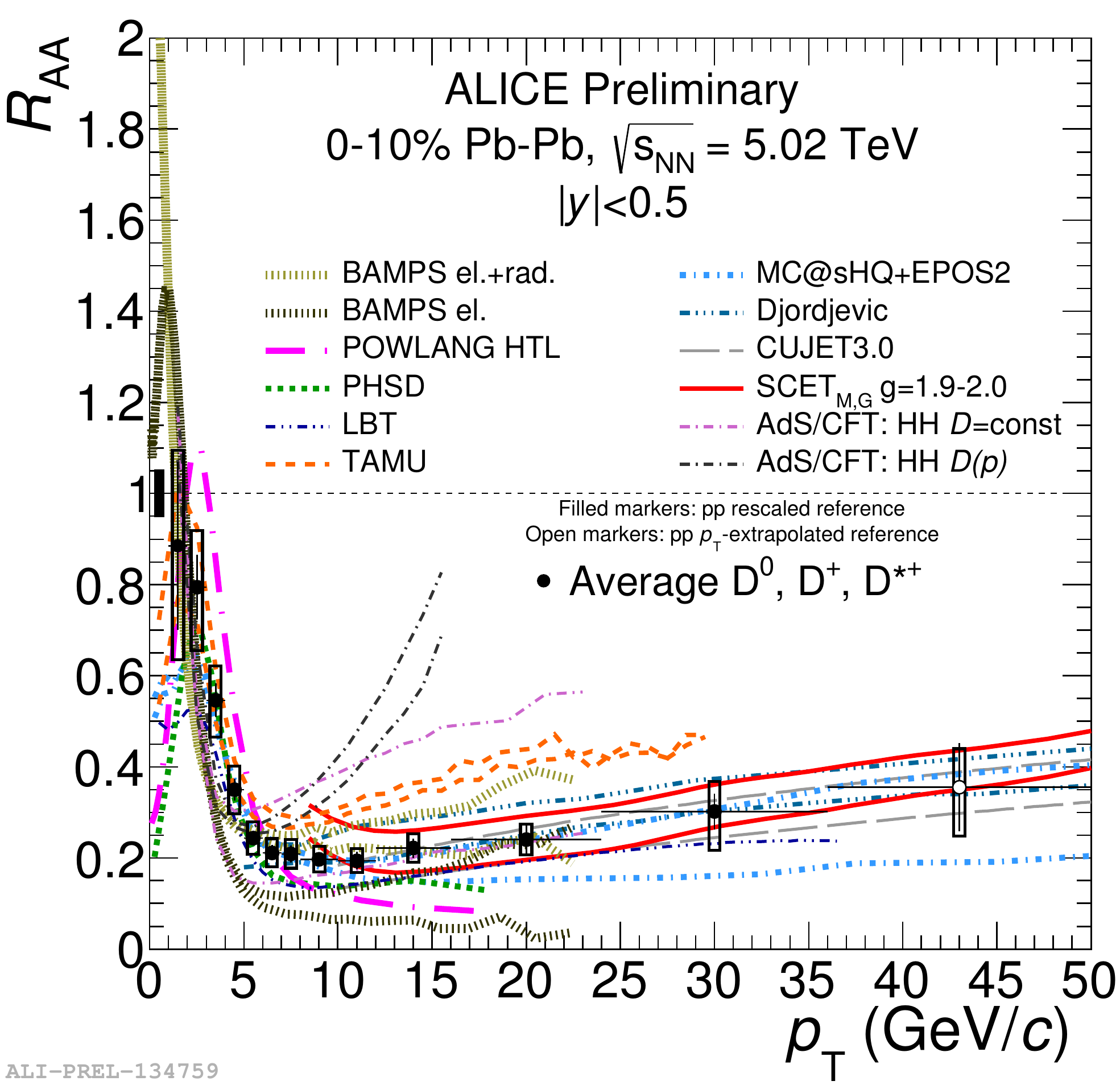} \hspace{0.5cm}
\includegraphics[width=6.5cm]{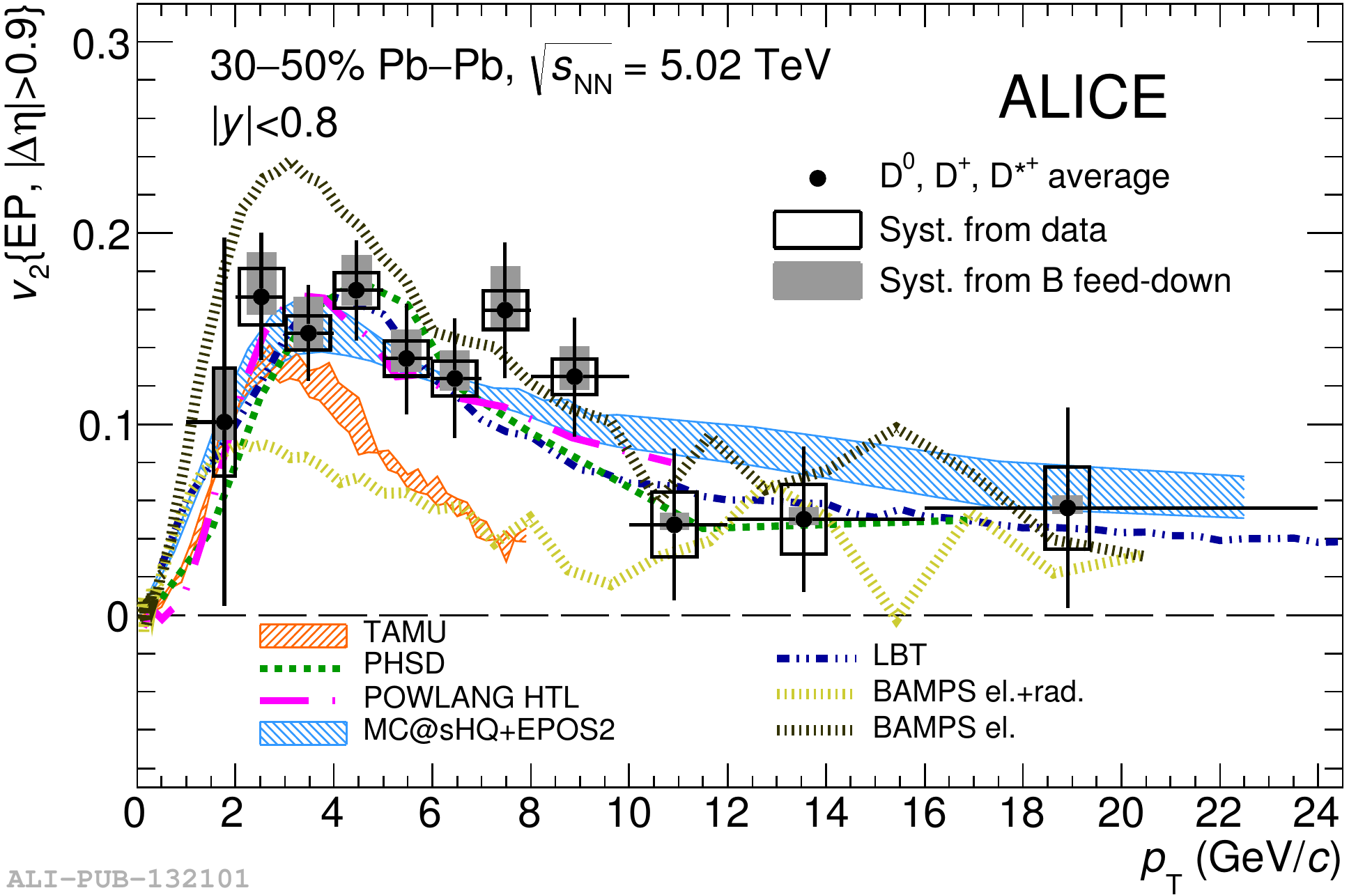}
}
\vspace*{8pt}
\caption{Average non-strange D-meson $R_{\text{AA}}$ in the 0--10\% centrality class (left)$^{15}$ and elliptic flow $v_{2}$ in 30--50\% centrality class (right)$^{19}$ measured in Pb--Pb collisions at $\sqrt{s_{\text{\tiny{NN}}}}$  = 5.02 TeV, compared with some of the available models.
\label{Fig3}}
\end{figure}

The average $R_{\text{AA}}$ of the three non-strange D-meson (D$^{0}$, D$^{+}$ and D*$^{+}$) in the 0--10\% centrality class (left) and $v_{2}$ in the 30--50\% centrality class (right) are compared with theoretical models in Fig. \ref{Fig3}. Models based on perturbative QCD calculation of parton energy loss \cite{pos}\cdash\cite{xu} provide a fair description of the $R_{\text{AA}}$ for $p_{\text{T}} >$ 10 GeV/$c$. All models (except of BAMPS and CUJET3.0) include a nuclear modification of the parton distribution functions as one of their ingredients. Models that rely on heavy-quark transport provide a fair description of the $R_{\text{AA}}$ at low $p_{\text T}$ and the D-meson $v_{2}$ as well.

\section{Conclusion}

The ALICE Collaboration measured the D mesons (D$^{0}$, D$^{+}$, D*$^{+}$ and D$^{+}_{\text s}$) $R_{\text{AA}}$ and the elliptic flow $v_{2}$ in Pb--Pb collisions at $\sqrt{s_{\text{\tiny{NN}}}}$  = 5.02 TeV. The comparison between the non-strange D-meson and D$^{+}_{\text s}$ gives a hint of charm hadronisation via coalescence. The results of the D meson elliptic flow $v_{2}$ show non-zero value for 2 $ < p_{\text T} < 10$ GeV/$c$ in 30--50\% centrality class, which provide information of the collective expansion of the system. The first measurement of the D$^{+}_{\text s}$ $v_{2}$ at the LHC has been reported. Event-Shape Engineering technique for the non-strange D-meson elliptic flow was applied.


\end{document}